\documentclass[preprint,aps,amsmath]{revtex4}
\usepackage{graphicx}
\usepackage{amssymb}
\usepackage{bm}

\newcommand\bnabla{\boldsymbol \nabla}

\begin{document}

\title{Interchange instability and transport\\in matter-antimatter plasmas}

\author{Alexander Kendl, Gregor Danler, Matthias Wiesenberger, Markus Held}
\affiliation{Institut f\"ur Ionenphysik und Angewandte Physik, Universit\"at
  Innsbruck, Technikerstr. 25, 6020 Innsbruck, Austria \vspace{1cm}} 

\begin{abstract}
\vspace{1cm}
Symmetric electron-positron plasmas in inhomogeneous magnetic fields are
intrinsically subject to interchange instability and transport. Scaling
relations for the propagation velocity of density ``blob'' perturbations
relevant to transport in isothermal magnetically confined electron-positron
plasmas are deduced, including damping effects when Debye lengths are large
compared to Larmor radii. The relations are verified by nonlinear full-F
gyrofluid computations. Results are in favour of sufficient magnetic confinement
for planned electron-positron plasma experiments. 
The model is generalised to other matter-antimatter plasmas. 
Magnetised electron-positron-proton-antiproton plasmas are susceptible to
interchange driven local matter-antimatter separation, which can be expected
to impede (so far unrealised) sustained laboratory magnetic confinement. 
\end{abstract} 

\maketitle

For reasons as yet elusive we are living in a world of ordinary matter only. 
Antimatter, predicted by Dirac less than a century ago \cite{dirac30}, has so
far made its earthly appearance through cosmic and laboratory high energy
events only in exiguous quantities \cite{alpha11} or rather brief periods
\cite{sarri15} before annihilating again. 
Macroscopic amounts of mixed matter-antimatter in the form of relativistic
electron-positron pair plasmas are presumed to exist at highest energy density
astrophysical objects, such as active galactic nuclei, pulsars
and black holes \cite{ruffini10}.

Extended confinement of larger numbers than a few antimatter particles in the
laboratory is impeded by instant annihilation at contact with materials. 
Ion and atom traps have been used to confine clouds of positrons
or antihydrogen \cite{danielson15,greaves97}. 
Investigation of many-body antimatter physics
on long time scales and low densities should be feasible in magnetically
confined quasineutral systems of charged particles
\cite{tsytovich78,pedersen03,pedersen12}.  Generation of high-density
matter-antimatter plasmas on short time scales could be achieved by
laser-target interaction \cite{sarri15}.

The most basic matter-antimatter system conceivable in the
laboratory is a classical plasma with equal amounts of electrons and positrons. 
An electron-positron (e-p) pair plasma is a unique many-body system and distinct from
a classical electron-ion (e-i) plasma in its intrinsic mass symmetry. 
Many fundamental instabilities that are abundant in mass-asymmetric e-i plasmas are
absent in corresponding e-p systems \cite{tsytovich78}. 
Micro-instabilities like drift waves in inhomogeneous magnetised e-i plasmas
generically result in turbulence and associated transport losses across the
confining magnetic field \cite{horton99}. 
The absence or weakness of instabilities and turbulence is
consequently crucial for the quality of planned magnetically confined e-p
plasma laboratory experiments \cite{pedersen03}. 

Classical resistive drift wave instabilities, which rely on asymmetry in the motion of
light electrons and heavier ions parallel to the magnetic field ${\bf B}$ with
electron to ion mass ratio $m_e / m_i \ll 1$, are non-existent in e-p plasmas. 
Magnetic curvature induced drift-interchange modes driven by an electron or positron
temperature gradient have been shown to be suppressed for e-p plasma densities
$N_e$ much smaller than the Brillouin density $N_B = \epsilon_0 B^2 / (2m_e)$
by an effective screening of electric potential fluctuations on scales smaller
than the Debye length \cite{pedersen03,helander14,helander16}. 
Cross-field pressure driven interchange motion and instability of plasma
perturbations is not only achievable in the presence of temperature gradients,
but also for isothermal plasmas in the presence of a density gradient. 
Local field-aligned density perturbations with a positive amplitude, often 
termed ``blobs'', can be pushed across the magnetic field (down the field
gradient) by their own intrinsic density inhomogeneity, even for constant
(flat) background density \cite{krasheninnikov01,dippolito11},
similar to the Rayleigh-Taylor fluid interchange instability in a
gravitational field \cite{garcia06}. 

Here we show that density blobs are unstable in an inhomogeneously magnetised
e-p plasma for a range of accessible parameters and thus can lead to
crucial transport losses. 
We present a simple relation to estimate the critical 
parameters for Debye stabilisation of e-p blob propagation, which we verify by
nonlinear full-F gyrofluid computations. 

In addition to e-p pair plasmas we also consider more general
matter-antimatter plasmas.
When the Debye length is much smaller than the drift scale, e-p plasmas
preserve quasi-neutrality and thus local equality of the electron and 
positron number densities $N_e ({\bf x},t) = N_p({\bf x},t)$ under the
influence of the drift-interchange instability. 
For quasi-neutral matter-antimatter ambi-plasmas consisting of initially equal
electron, proton, positron and antiproton densities (which could in some
future experiment be feasible, but is cosmologically probably irrelevant) 
we show that such a (quasi-neutral) one-to-one correspondence of matter and
antimatter particle densities is no longer maintained: the ambi-plasma
interchange instability does not only lead to transport but also to spatial
matter-antimatter separation.

First, we analyse magnetised e-p plasmas by means of a full-F gyrofluid
model \cite{madsen13}, which is derived from a gyrokinetic model that evolves
the full distribution function $F({\bf x}, {\bf v}, t)$. 
In the isothermal two-dimensional limit \cite{wiesenberger14, kendl15} the
model consists of continuity equations for the gyrocenter densities $n_s$ 
for electron and positron species $s \in (e, p)$, 
\begin{equation} 
\partial_t n_s =   {1 \over B} \left[n_s, \psi_s \right] 
 + n_s {\hat \kappa} (\psi_s) + \tau_s {\hat \kappa} (n_s), \label{eq:den0}
\end{equation}
coupled by polarisation to the electric potential $\phi$ in Poisson's equation:
\begin{equation} 
\sum_s Z_s N_s + \varepsilon \; \nabla_{\perp}^2 \phi = 0.
\label{eq:pol0}
\end{equation}

Exact local quasi-neutrality is thus violated for $\varepsilon \neq 0$.
The particle densities $N_s$ with charge states $Z_e=- 1$, $Z_p=+1$ are linked
to the gyrocenter densities $n_s$ via
\begin{equation} 
N_s = \Gamma_{1s} n_s + \bnabla \cdot \left( n_s  {\mu_s \over Z_s
    B^2} \bnabla_{\perp} \phi \right).
\label{eq:gyrocenter}
\end{equation}

In contrast to strictly quasineutral e-i gyrofluid systems as in 
ref.~\cite{madsen13}, we do not invoke the gyrokinetic ordering ($\varepsilon
\ll 1$), but retain in our present e-p model the Debye parameter 
\begin{equation} 
\varepsilon = \left( {\lambda \over \rho} \right)^2 = {2N_B \over N_{e0} },
\end{equation}
which represents effects of finite Debye length
$\lambda = \sqrt{\epsilon_0 T_e/(e^2 N_{e0})}$ in relation to the Larmor radius
$\rho=\sqrt{T_e m_e}/(eB_0)$.   Here $T_e$ is the electron temperature, $m_e$
the electron mass, and $B_0$ a reference magnetic field strength.
The temperature ratio is $\tau_s = T_s / (Z_s T_e)$ so that $\tau_e = -1$ and
$\tau_p \geq 0$. 
The e-p mass ratios  $\mu_s = m_s / m_p = 1$ are both unity so that 
$\mu \equiv \sum_s \mu_s = \mu_e + \mu_p =2$, in contrast to e-i plasmas where 
$\mu = \mu_e + \mu_i \approx \mu_i=1$.
Annihilation between positrons and electrons is here neglected for the 
time scales of blob propagation \cite{helander03}.

Finite Larmor radius and ponderomotive effects enter via 
$\psi_s = \Gamma_{1s} \phi - (1/2) (\mu_s / Z_s B^2) (\nabla_{\perp} \phi)^2$.
The gyro-averaging operator in Pad\'e approximation is defined by
$\Gamma_{1s} = (1 + (1/2) b_s)^{-1}$  with $b_s = - (\mu_s \tau_s / Z_s)
\nabla_{\perp}^2 = k_{\perp}^2$ here. 
The gyrocenter densities $n_s$ have been normalised to a constant reference
density $n_0$, so that the magnitude of the plasma density $n_s \leftarrow
n_s/n_0$  is of order one, and the potential is normalised to $\phi \leftarrow
(e \phi/T_e)$. 
Perpendicular scales and spatial derivative operators are normalised
to the drift scale $\rho$ so that $\nabla \leftarrow \rho \nabla$.
The time scale is normalised as $\partial_t \leftarrow (\rho / c_s)
\partial_t$ with sound speed  $c_s = \sqrt{T_e/m_e}$. 
The 2-d advection terms are expressed through Poisson brackets
$[f,g] = (\partial_x f)(\partial_y g) - (\partial_y f)(\partial_x g)$ 
for local coordinates $x$ and $y$ perpendicular to ${\bf e}_z = {\bf B}/|B|$.   
Normal magnetic curvature $\kappa = \partial_x \ln B \equiv 2\rho/R$ 
(including magnetic curvature and gradient drifts) 
enters into $\hat \kappa = \kappa \; \partial_y $.
For most toroidal magnetic confinement experiments the gyro scale $\rho$ is
much smaller than the effective curvature radius $R$, which can be well
approximated by the torus radius, so that $\kappa \ll 1$.

We analyse the dependence of the ideal interchange growth rate on
the Debye screening parameter $\varepsilon$ in the present model 
by linearising eqs.~(\ref{eq:den0}) and (\ref{eq:pol0}) 
on top of a background density $n_0({\bf x})$
for small perturbation amplitudes proportional to $\tilde n_e, \tilde n_p,
\tilde \phi \sim \exp{(-i\omega t+i{\bf k} \cdot {\bf x})}$ and small curvature.
For simplicity we neglect finite Larmor radius effects in the analytical
treatment and set $b_s=0$.
We get the linear system of equations, with $s\in(e,p)$:
\begin{eqnarray}
-\omega \tilde n_s &=& (\omega_c - \omega_{\ast}) \tilde \phi + \tau_s
\omega_c \tilde n_s, \\
(\mu + \varepsilon) k_{\perp}^2 \tilde \phi &=& \tilde n_p - \tilde n_e.
\label{eq:lin}
\end{eqnarray}
Here $\omega_c = \kappa k_y$ and $\omega_{\ast} = - g k_y$, with 
$g = \partial_x \ln n = - L_n^{-1}$ for a constant density gradient
length $L_n$, where $g=g_0+g_1$ is composed of a background gradient $g_0$
and a contribution $g_1$ by the intrinsic blob front.
The linearisation of the Poisson equation resembles a Boussinesq approximation. 
The system is easily resolved into a dispersion relation $\omega=\omega({\bf k},\varepsilon)$.
For $\omega = \omega_r + i \gamma$ we obtain an interchange growth rate
$\gamma$ with
\begin{equation} 
\gamma^2 = { 1+\tau \over \mu + \epsilon }
\left[ {\omega_c \over k_{\perp}^2} (\omega_{\ast} - \omega_c) -{1 \over 4} (\mu
  + \epsilon) (1 + \tau) \omega_c^2 \right].
\label{eq:gamma}
\end{equation} 
The model is interchange unstable for $\gamma^2>0$. Here for e-p plasmas
$\mu=2$ (with $\tau \equiv \tau_p$ and $\tau_e = -1$), 
whereas for quasineutral e-i plasmas we would have $\mu=1$ and $\epsilon = 0$.

We apply the blob correspondence principle \cite{krasheninnikov01} 
which assumes that the perturbation mode most relevant for instability of the
blob  front and thus for the resulting blob propagation is of the actual
initial blob scale $\sigma$.
For Gaussian blobs with width $\sigma$, which relates $k_y = k_x \approx 1/
\sigma$ and $k_{\perp}^2 = k_x^2 + k_y^2 \approx 2 / \sigma^2$. 
Relating the growth rate to blob convection by 
$i \gamma \sim (d/dt) \sim v_x \partial_x \sim i v_x / \sigma$, 
the blob propagation velocity can be approximated as $v_x = \sigma \gamma$. 

We here neglect a background gradient so that $g_0=0$, and $g_1 = \partial_x
\ln n \approx (1/n_0) \partial_x n$ is evaluated for an initial Gaussian
blob density $n({\bf x}) = n_0 + A \exp(-{\bf x}^2/\sigma^2)$ at the location
$x=\sigma/\sqrt{2}$ of the steepest front gradient to be  
$g_1 = - \sqrt{(2/e)} \cdot A/\sigma \equiv - a/\sigma$ with $a \approx 0.86 A$.

In the following we set $\tau=\tau_p=1$, $\mu=2$.
In the blob correspondence principle approximation we thus obtain
\begin{equation}
v_{x} = \frac{1}{\sqrt{1+ (\varepsilon / 2)}} \sqrt{ \frac{1}{2} a \kappa
    \sigma  - \kappa^2 \sigma^2 \left[ \frac{1}{2} + 
      \frac{1+ (\varepsilon /2)}{\sigma^2} \right] }.  
\label{eq:vxfull}
\end{equation}

\begin{figure} 
\vspace{0.5cm}
\includegraphics[width=11.0cm]{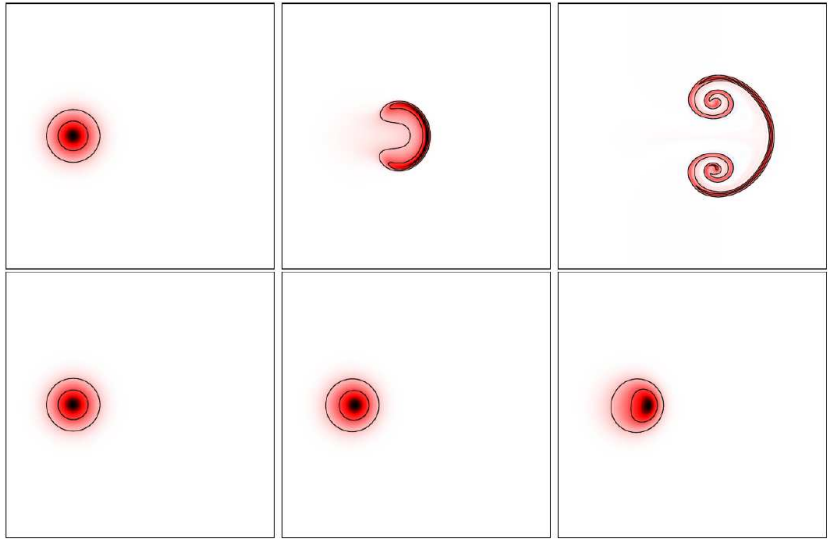}
\caption{\sl Interchange driven propagation of an initially Gaussian density
  perturbation 
  in an inhomogeneous magnetic field: comparison between exactly quasi-neutral
plasma with $\varepsilon=0$ (top row) to Debye screened plasma with
$\varepsilon=(\lambda/\rho)^2 = 50$ (bottom row), shown at three times (from
left to right: $t=0$, $t=25000$, $t=50000$ in units of $\rho/c_s$).
The colour scale indicates positive blob density compared to the background
($n_0=1$), with contour lines drawn at $n=1.1$ and $n=1.3$.}  
\label{fig-struct}
\end{figure}

For $\kappa \sigma \ll 1$ the second term can be neglected (except for very
large $\varepsilon$ or very small amplitudes) so that the e-p blob velocity
can be approximated as
\begin{equation}
v_{x0} (\varepsilon) \approx \frac{v_0}{\sqrt{1+(\varepsilon/2)}}  \quad
\mbox{with} \quad v_0 = \sqrt{ 0.43 \; A \kappa \sigma}.
\label{eq:vxapprox}
\end{equation}
We verify the (range of) validity for the relations in
eqs.~(\ref{eq:vxfull}) and (\ref{eq:vxapprox}) by numerical simulation of blob
propagation for various parameters in the full-F nonlinear model
eqs.~(\ref{eq:den0}, \ref{eq:pol0}) using the codes TOEFL \cite{kendl14,kendl15} and FELTOR \cite{wiesenberger14,danler15}.   
During its propagation down the magnetic field gradient the unstable blob changes 
its shape in the simulations from initially circular O $\rightarrow$ D
$\rightarrow$ $\supset$ to mushroom cap shape, with subsequent vortex roll-up
and final turbulent break-up. 
Typical density structures of the rightward propagating blob at three different times
for $\varepsilon=0$ (top) and $\varepsilon=50$ (bottom) are shown in
Fig.~\ref{fig-struct} for an amplitude $A=0.5$, width $\sigma=10$ in units
of $\rho$, and curvature $\kappa=10^{-4}$. 
The computational domain is (128 $\rho$)$^2$ with a grid resolution of $512^2$.
The simulation illustrates that blob transport is for these parameters
strongly inhibited by Debye screening with $\varepsilon=50$ (bottom).

We compare the analytical propagation velocity
estimates $v_x$ with the maximum center-of-mass velocity 
$v_{sim} \equiv \mbox{max}(v_{com})$ in the blob simulations, 
where $v_{com} \equiv {\dot x_{com}}$ with 
$x_{com} \equiv [\int dx dy(n_e(x,y)-n_0)x]/[(\int dx dy (n_e(x,y)-n_0)]$ for
mass symmetric e-p plasmas.
The maximum velocity, which coincides for gyrocenter and particle densities,
is usually obtained in the bean (``D'') shaped phase.

\begin{figure} 
\vspace{0.5cm}
\includegraphics[width=9.cm]{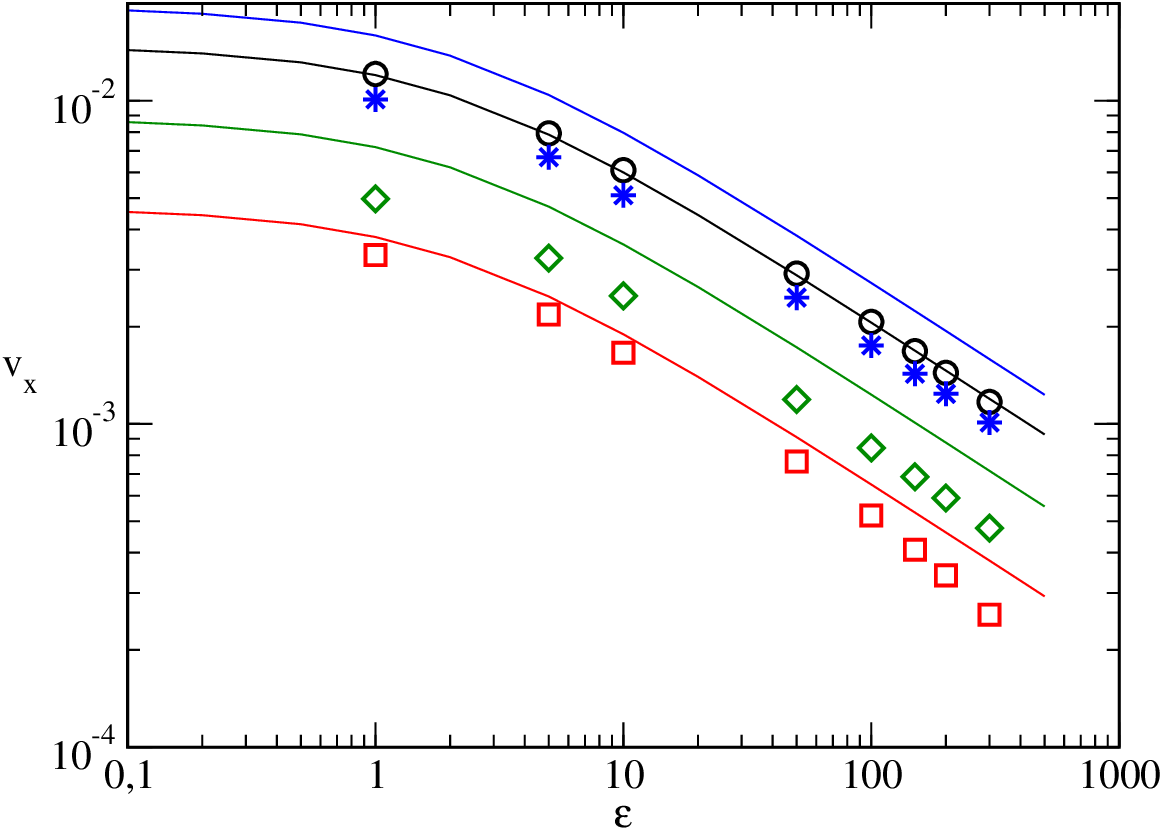} 
\caption{\sl Scaling of simulated maximum electron-positron blob velocities
  $v_{sim}$ (symbols) and theoretical velocity scaling $v_x$ (curves) with Debye
  screening parameter $\varepsilon$.  
  For all blob  parameters the velocities qualitatively follow the scaling 
  $v_{x0} (\varepsilon) = v_0  / \sqrt{1+(\varepsilon/2)}$ over several orders
  of magnitude.   
  Black/circles: $A=0.5$, $\sigma=10$; blue/stars: $A=0.1$, $\sigma=100$;
  green/diamonds: $A=0.05$, $\sigma=40$; red/squares: $A=0.05$, $\sigma=10$.
  Quantitative agreement of simulations with the analytical scaling law
  is achieved only for large blob amplitudes and/or small blob widths
  (e.g. here the case shown in black, less for red).} 
\label{fig-eps}
\end{figure}

In Fig.~\ref{fig-eps}, $v_{sim}(\varepsilon)$ computed from the full model
simulation (symbols), and the corresponding analytical estimates
$v_x(\varepsilon)$ from eq.~(\ref{eq:vxapprox}) (continuous curves) are
plotted as a function of Debye screening $\varepsilon$ for 
various blob parameters.
In all cases the velocities qualitatively correspond well to the scaling 
$v_{x0} (\varepsilon) = v_0  / \sqrt{1+(\varepsilon/2)} $
over several orders of magnitude.   
The four cases shown here are: $A=0.5$,  $\sigma=10$ (black/circles);
$A=0.1$, $\sigma=100$ (blue/stars); $A=0.05$, $\sigma=40$ (green/diamonds); 
$A=0.05$, $\sigma=10$ (red/squares). 
The curvature parameter is $\kappa = 10^{-4}$.
Simulation domain sizes are adapted to ensure at least $L_y > 10 \sigma$ in
order to minimise boundary effects.
Quantitative agreement between the linear scaling law (curves) and
simulations is best achieved for large blob amplitudes and/or small blob
widths (i.e. the cases shown in red and black).
The largest quantitative differences between analytical estimate and
simulations appear for the cases with small amplitudes and large blob sizes
(shown here in blue and green), which is in agreement 
with an e-i scaling law based on energetic principles \cite{kube16}.
A noticeable qualitative deviation is observed in the (red) case with both
small amplitude and small blob width for large $\varepsilon$: in connection
with the slow propagation velocity (and longer simulation time), numerical
viscosity and diffusion effects can in this case play a larger role. 
The blob propagation velocities drop by more than an order of magnitude for
$\varepsilon > 200$. 

In experiments with magnetically confined e-p plasmas the interchange
driven transport of blob-like random density fluctuations is consequently
effectively reduced if large values of $\varepsilon$ can be achieved.
For the projected APEX electron-positron stellarator experiment
\cite{pedersen12} the major torus radius is $R=15$~cm, and planned plasma
temperatures are in the order 
of 0.2 - 2 eV at a magnetic field strength of $B=2$~T. This results in an
electron/positron gyro radius of $\rho \sim 10^{-3}$~cm. 
Foreseen particle densities are around $10^{13}$~m$^{-3}$, which results in
Debye lengths $\lambda \sim 0.1-0.3$~cm \cite{pedersen12}.
The curvature parameter thus is in the order of $\kappa \sim 10^{-4}$ and the
Debye parameter is expected around $\varepsilon \sim 50 - 300$. 
An unknown quantity in the estimations of interchange driven transport is the
size and amplitude of appearing density perturbations. In e-i tokamak edge plasmas
the observed blob structures are generated by drift wave vortices or are sheared off by
perpendicular flows, which results in typical sizes in the order of a few to
20 (ion) drift scales. Blob and vortex amplitudes in the tokamak edge
are found in the range around $A \sim 0.01 - 1$ relative to background densities.
For magnetically confined e-p plasmas we can not assume the same parameters for
initial seed perturbations, as driving by drift wave vortices is absent, and the
possible flow shearing rate is unknown. But as the e-p plasma is not
exactly quasi-neutral, small local electric field or density perturbations can
appear spontaneously with cross-field extensions up to the order of the Debye length. 
It therefore appears appropriate to consider blob sizes in the order between
$\sigma \sim 1-200$, with a wide range of possible initial perturbation
amplitudes (as for the computations in Fig.~\ref{fig-eps}).
We stress that an additional background density gradient (which is expected
in any magnetically confined plasma) will add to the interchange driving rate.

We now discuss parameters required for complete stabilisation of e-p blobs.
As can be seen from eqs.~(\ref{eq:gamma}) and (\ref{eq:vxfull}),
the system is stable for $\varepsilon=0$ when 
$(0.43 A \kappa \sigma - 0.5 \kappa^2 \sigma^2 - \kappa^2) < 0$. 
This is achieved for blob amplitudes less than $A_{crit} = 1.16 \kappa \sigma$. 
Here we are more interested in the stability criterion for finite Debye
parameter $\varepsilon$, 
which leads to a minimum  $\varepsilon$ for stabilisation:
\begin{equation}
 \varepsilon \geq \sigma \left( \frac{a}{\kappa} - \sigma \right) -2 \approx
 \sigma^2 \left( \frac{a}{\kappa \sigma} - 1 \right). 
\label{eq:mindebye}
\end{equation}
For $\sigma=10$, $\kappa =  10^{-4}$ and $A=0.01$ this corresponds to
$\varepsilon_{crit} \approx 700$. For larger amplitudes the limit is higher.  
This blob stabilisation is also observed in simulations. 
For example, $\varepsilon \approx 300$ is expected in the planned APEX
experiment \cite{pedersen12}. This would not be sufficient to completely
stabilise self-propelled interchange blobs of sizes in a wide range of relevant
parameters by Debye screening, but particle transport $\Gamma \sim A v_x \sim
1/\sqrt{1+(\varepsilon/2)}$ would nevertheless be significantly reduced by around an 
order of magnitude (compared to $\varepsilon = 0$).
In order to maintain large $\varepsilon = (\lambda/\rho)^2 = (2 N_B / N_{e0})$
in e-p experiments, the particle densities have to be kept at sufficiently
small values $N_{e0} \ll N_B = \epsilon_0 B^2/(2m_e)$.

In addition to the self-propelled blobs which require finite amplitude, 
the linear analysis also shows interchange instability for arbitrary small
perturbations on an additional large enough background density gradient. 
But as both the perturbation amplitudes and widths to be expected in an e-p
plasma are largely unknown, prediction of the actual quantitative level of
convective interchange driven transport in toroidal magnetically confined e-p
plasmas is difficult. 
As an order of magnitude estimate for cross-field transport in the
APEX experiment ($N \sim 10^{13}$~ m$^{-3}$), we arrive at a radial density
flux $\Gamma = (A N) (v_x c_s) \approx (0.01 \cdot 10^{13}$~m$^{-3}) \cdot
(0.01 \cdot 10^5$~m~s$^{-1}) \approx 10^{14}$~m$^{-2}$~s$^{-1}$. 
For an APEX scale plasma surface $S\sim 0.1$~m$^2$ this would correspond
to a loss rate of $10^{13}$ particles per second 
and a consequent confinement time of milliseconds to seconds, depending
crucially on the possible fluctuation amplitudes and scales, which should (in
the absence of other instabilities) be given by the thermal background
fluctuation level. Our results on self-propelled or density gradient driven
blob transport complement the discussion of temperature gradient driven
instabilities in e-p plasmas in refs.~\cite{pedersen03,helander14,helander16}
and to some extent confirm the optimistic confinement expectations for the
APEX experiment \cite{pedersen12}.  

The apparent prospect of creating macroscopic magnetically confined
electron-positron plasmas in the laboratory motivates to consider other
possible many-body matter-antimatter systems. 
From symmetry principles, it appears attractive to study a quasi-neutral
system consisting of equal numbers of electrons, protons, positrons and
antiprotons. It is unclear and, regarding present standard models of baryogenesis
and magnetogenesis, highly unlikely, whether such a magnetised ``ambi-plasma''
(following a terminology introduced by Hannes Alfv\'en) has actually ever
existed in cosmological history, but it might in principle be created in
laboratory when large numbers of positrons and antiprotons could be supplied
continuously at one site (cf.~ref.~\cite{alpha11}).

We therefore briefly complement the discussion of our results presented above
by computations of quasi-neutral density blob perturbations in such an ambi-plasma.
The model is again given by eqs.~(\ref{eq:den0}), (\ref{eq:pol0}) and
(\ref{eq:gyrocenter}) with $s \in (e,i,p,a)$ consisting of electrons, ions
(protons), positrons and antiprotons. 
All scales are now normalised to the proton drift scale 
$\rho=\sqrt{T_e m_i}/(eB)$, and Debye screening is neglected ($\varepsilon=0$).
We initialise with a matter-antimatter symmetric Gaussian density perturbation
with same parameters $n_s({\bf x}) = n_0 + A \exp(-{\bf x}^2/\sigma^2)$ for
all $s \in (e,i,p,a)$. FLR effects are now neglected for the leptons (e, p),
but retained for the baryons (i, a) in $\tau_i = 1$ and $\tau_a = -1$. 
The overall evolution of the ambi-plasma blob is at the beginning similar to
the case of e-i or e-p blobs: the magnetic curvature induced interchange
mechanism generates an electric dipole potential $\phi({\bf x})$ with an 
associated vorticity field $\Omega({\bf x}) = \nabla^2 \phi$ that
drives the blob down the magnetic field gradient.

\begin{figure} 
\includegraphics[width=7.0cm]{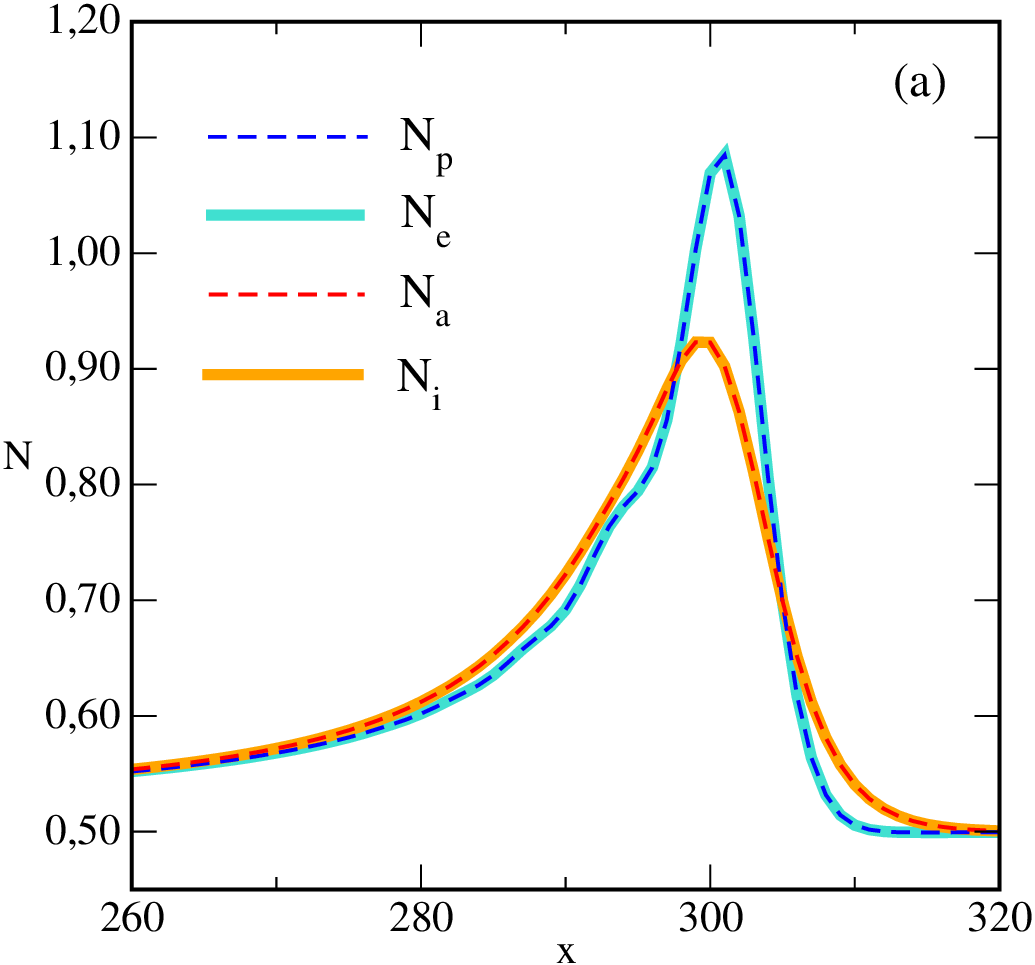}
\includegraphics[width=7.0cm]{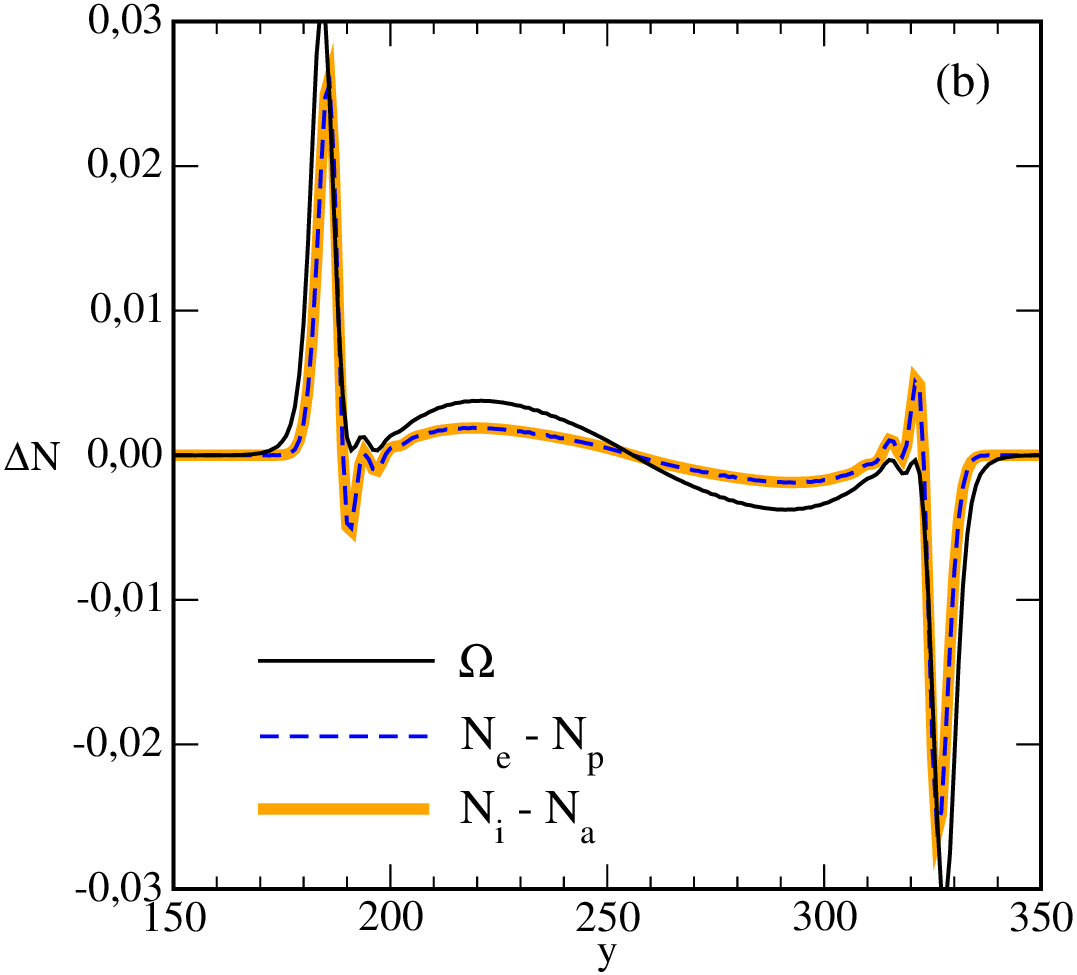}
\caption{\sl Species asymmetry in an ambi-plasma blob. 
(a) Cut in propagation ($x$) direction through blob center: electron and
  positron particle densities ($N_e$, $N_p$) deviate from proton and antiproton
  densities ($N_i$, $N_a$). (b) Cut in  perpendicular ($y$) direction through
  blob center: differences in particle-antiparticle densities are
  correlated with the vorticity $\Omega$.
Cross-sections are plotted for a mushroom cap blob shape corresponding to the
top right frame in Fig.~1.}    
\label{fig-ambi}
\end{figure}

Remarkably, the detailed spatial matter-antimatter symmetry is partially
broken during 
the further blob development in two ways: Fig.~\ref{fig-ambi} (a) shows that the
particle densities $N_s(x)$ clearly start to differ between species. 
The electron and positron densities are closely aligned with each other, and
so, respectively, are the proton and antiproton densities:  
leptonic and baryonic densities locally deviate from each other.

Fig.~\ref{fig-ambi} (b) reveales that also the particle-antiparticle symmetry is
locally broken, where both $(N_e-N_p)$ and $(N_i-N_a)$ differences grow with
time. The curves for the electron-positron density difference (dashed blue)
exactly agree with the proton-antiproton density difference (bold orange)
because of the quasineutrality condition ($\sum_s Z_s N_s=0$).
Both appear strongly correlated with the vorticity $\Omega$.

The local baryon-lepton asymmetry in particle densities $N_s$ is caused by
finite Larmor radius (FLR) effects, which mainly enter via 
$N_s \sim \Gamma_{1s} n_s = n_s/(1+(1/2) b_s)$ and reduce the particle
densities of the species (baryons) with relevant masses $\mu_s$ in regions of
steep gradients. This asymmetry disappears when FLR effects are switched off
in the simulation. 

The breaking of the local particle-antiparticle symmetry is, in contrast, a
result of alignment with vorticity \cite{kendl12} driven by the interchange
instability, and consequently grows with time during the blob
evolution.  The asymmetry is again most pronounced at steepening blob edges.
It persists when $\tau_i = \tau_a = 0$ and is thus not related to FLR effects.
Taking the total time derivative of the polarisation equation (linearised,
without FLR terms) we have 
$D_t (N_p- N_e) = D_t (N_i - N_a) \sim {\hat \kappa} (N_e + N_p) \sim 
(\mu_i + \mu_a) D_t \Omega$ with $\Omega = \nabla_{\perp}^2 \phi$.

A consequence of the particle-antiparticle asymmetry appears particularly in
the presence of annihilation, which can play a stronger role on the ion blob
evolution time scale (that is larger by $m_i/m_e$ compared to
e-p blob time scales), and will enhance the relative particle-antiparticle
density difference. As the interchange drive is present in any
inhomogeneously magnetized plasma, matter-antimatter ambi-plasmas will loose
their symmetry on combined interchange and annihilation time scales, and in
the worst case will result in patches of electron-proton matter locally
separated from positron-antiproton antimatter plasma.
This matter-antimatter separation by plasma instabilities reduces the prospect
of achieving magnetic confinement of laboratory ambi-plasma.

In summary, we have shown that transport of plasma density by interchange
driving appears for a large range of parameters in e-p plasmas, but can
(similarly to previously discussed temperature gradient driven modes) be
significantly reduced by maintaining a high enough density to facilitate
damping by Debye screening.  
Magnetic confinement of e-p plasmas for relevant time scales so appears to be
feasible. In contrast, we have further argued that long confinement of
matter-antimatter ambi-plasmas consisting of globally equal numbers of
electrons, protons, positrons and antiprotons is likely inhibited by local
matter-antimatter separation in the presence of annihilation.


\newpage

{\sl Acknowledgement:}
This work was supported by the Austrian Science Fund (FWF) project Y398.
The computational results presented have been achieved in part using the
Vienna Scientific Cluster (VSC).

\end{document}